\documentclass[11pt,a4paper]{article}
\pdfoutput=1
\usepackage{jheppub}

\usepackage{amsmath,colonequals}
\usepackage{amssymb}
\usepackage{graphicx}
\usepackage{epsfig}
\usepackage{bm}
\usepackage{color}

\newfont{\sss}{cmmi10 at 20.74pt}

\def\spose#1{\hbox to 0pt{#1\hss}}
\def\simlt{\mathrel{\spose{\lower 3pt\hbox{$\mathchar"218$}}
     \raise 2.0pt\hbox{$\mathchar"13C$}}}
\def\simgt{\mathrel{\spose{\lower 3pt\hbox{$\mathchar"218$}}
     \raise 2.0pt\hbox{$\mathchar"13E$}}}
\def\0{\mbox{\boldmath$\displaystyle\mathbf{0}$}}

\def\be{\begin{equation}}
\def\ee{\end{equation}}
\def\bea{\begin{eqnarray}}
\def\eea{\end{eqnarray}}

\newcommand{\ce}{\colonequals}

\title{CP violatingTri-bimaximal-Cabibbo mixing}

\author{D. V. Ahluwalia}

\affiliation{Department of Physics and Astronomy,
Rutherford Building, University of Canterbury,
Private Bag 4800, Christchurch 8140, New Zealand}

\emailAdd{dharamvir.ahluwalia@canterbury.ac.nz}


\abstract{ In view of the new data from the Daya Bay and RENO collaborations, King has presented a very natural deformation of tri-bimaximal mixing. Here we show that $L/E$ flatness of the $e$-like event ratio in the atmospheric neutrino data, when coupled with King's observation that the smallest neutrino mixing angle, $\theta_{13}$, seems to be related to the largest 
quark mixing angle (the Cabibbo angle $\theta_C$), leads to a CP violating  tri-bimaximal-Cabibbo mixing. King's tri-bimaximal-Cabibbo mixing follows as a leading order approximation from our result.  }

\notoc
\begin{document}

\maketitle
\bigskip\bigskip

\hrule
The precise form of the neutrino mixing matrix, $U$,
that defines the relationship between the flavour and mass eigenstates, $\vert\nu_\ell\rangle$ and $\vert\nu_j\rangle$ respectively~\cite{Chau:1984fp,Beringer:2012bj}, reads 
\begin{equation}
\vert\nu_\ell\rangle = \sum_j U^\ast_{\ell j} \vert\nu_j\rangle, \quad\ell = e,\mu,\tau,\quad j=1,2,3,
\end{equation}
and  the knowledge of the masses for the underlying mass eigenstates, arise from yet unknown physics.  Nevertheless, once the parameters that determine the mixing matrix and the mass-squared differences are deciphered from the data one can derive their phenomenological consequences on supernova explosions~\cite{Ahluwalia:2004dv,Lunardini:2007vn,Duan:2006an,Duan:2007sh},  on the synthesis of elements~\cite{Yoshida:2006sk},  on the cosmic microwave background and the distribution of large-scale structure~\cite{Lesgourgues:2006nd}. In particular, if the neutrino mixing angle $\theta_{13} \ne 0$ then one can obtain CP violation in the neutrino sector with many interesting physical consequences~\cite{Khlopov:1981nq,Frampton:2002qc,Balantekin:2007es}.

The T2K, MINOS, and Double CHOOZ indications that the smallest neutrino mixing angle $\theta_{13}$ may  be non-zero~\cite{Abe:2011ph,Adamson:2011qu,Abe:2011fz} has now been confirmed by the results of the Daya Bay and RENO collaborations~\cite{An:2012eh,Ahn:2012nd}. King has  made the observation~\cite{King:2012vj} that the smallest neutrino mixing angle $\theta_{13}$, seems to be related to the largest 
quark mixing angle, the Cabibbo angle $\theta_C$~\cite{Cabibbo:1963yz}, or equivalently to the Wolfenstein parameter, $\lambda = 0.2253\pm0.0007$~\cite{Wolfenstein:1983yz,Beringer:2012bj}:\footnote{It is worth noting that Mohapatra and Smirnov had earlier conjectured King's observation~\cite[Sec.~3.1]{Mohapatra:2006gs}.}
\begin{equation}
 \theta_{13} ~ \mbox{(or, } \theta_{reac}\mbox{)} = \arcsin\left(\frac{\sin\theta_C}{\sqrt{2}}\right) = \arcsin\left(\frac{\lambda}{\sqrt{2}}\right).
\end{equation}
To this observation we now add that the $L/E$ \textemdash~where $L$ is the neutrino source-detector distance and $E$ is the neutrino energy ~\textemdash ~flatness of the $e$-like event ratio observed for atmospheric neutrinos~\cite{Fukuda:1998mi} requires that
\begin{equation}
\theta_{23}~ \mbox{(or, } \theta_{atm}\mbox{)}  = \frac{\pi}{4}, \quad \delta= \pm \frac{\pi}{2}.\label{eq:2}
\end{equation}
This observation was first made in reference~\cite{Ahluwalia:2002tr}.  The $\delta$ obtained in~\cite{Ahluwalia:2002tr} was also introduced recently as an Ansatz in Ref.~\cite{Zhang:2012ys}.

Global analysis of neutrino oscillation data by two independent groups shows: (a) $\delta$ to be $\left(0.83^{+0.54}_{-0.64}\right)\pi$ for the normal mass hierarchy while allowing for the full $[0,2 \pi]$ range for the inverted mass hierarchy~\cite{Tortola:2012te}, (b) $\delta \approx \pi$ with no significant difference between the normal and inverted mass hierarchies~\cite{Fogli:2012ua}. A detailed study of these two papers reveals that there is no statistically significant indication which disfavours $\delta  = \pm \pi/2$. Regarding $\theta_{23}$: (a)  the first of the mentioned groups obtains $\sin^2\theta_{23} = 0.49^{+0.08}_{-0.05}$ for the normal mass hierarchy, and $\sin^2\theta_{23} = 0.53^{+0.05}_{-0.07}$ for the inverted mass hierarchy (these values are consistent with $\theta_{23}=\pi/4$), while (b) the second group finds a slight preference for $\theta_{23} < \pi/4$.

Both groups agree with the tri-bimaximal mixing value for the remaining angle~\cite{Tortola:2012te,Fogli:2012ua}
\begin{equation}
\theta_{12}~ \mbox{(or, } \theta_{\odot}\mbox{)}  =  \arcsin\left(\frac{1}{\sqrt{3}}\right).
\end{equation}
With all the angles and phases thus fixed, the neutrino mixing matrix for the choice $\delta  = \pi/2$ in equation~(\ref{eq:2}) takes the form

\begin{equation}
U^+ = \begin{pmatrix}
\sqrt{\frac{2}{3}}
\left(1-\frac{\lambda^2}{2}\right)^{1/2} & 
\sqrt{\frac{1}{3}} \left(1-\frac{\lambda^2}{2}\right)^{1/2} & i \frac{1}{\sqrt{2}}\lambda\\ 
-\frac{1}{\sqrt{6}} \left(1- i \lambda\right) &
\frac{1}{\sqrt{3}}\left(1 + i \frac{1}{2}\lambda\right) &
\frac{1}{\sqrt{2}}\left( 1-\frac{\lambda^2}{2} \right)^{1/2}\\
\frac{1}{\sqrt{6}} \left(1+ i \lambda\right) &
-\frac{1}{\sqrt{3}}\left(1 - i \frac{1}{2}\lambda\right) &
\frac{1}{\sqrt{2}}\left( 1-\frac{\lambda^2}{2} \right)^{1/2}
\end{pmatrix}.
\end{equation}
Its counterpart, $U^-$, for $\delta = - \pi/2$ is obtained by letting $i\to -i$ in $U^+$. As a measure of CP violation, following~\cite{Beringer:2012bj}, we define the asymmetries
\begin{equation}
A_{CP}^{(\ell^\prime\ell)}\ce P(\nu_\ell \to \nu_{\ell^\prime}) -
 P(\bar\nu_\ell \to \bar\nu_{\ell^\prime}),
\end{equation}
and find 
\begin{align}
A_{CP}^{(\mu e)} = - A^{(\tau e)}_{CP} = A_{CP}^{(\tau\mu)} & =  \mp\frac{1}{3} \lambda\left(2 - \lambda^2\right) \left(
\sin \frac{\Delta m^2_{32}}{ 2 p} L
+ \sin \frac{\Delta m^2_{21}}{ 2 p} L
+ \sin \frac{\Delta m^2_{13}}{ 2 p} L
\right)\nonumber\\
&\approx \mp0.146 \left(
\sin \frac{\Delta m^2_{32}}{ 2 p} L
+ \sin \frac{\Delta m^2_{21}}{ 2 p} L
+ \sin \frac{\Delta m^2_{13}}{ 2 p} L
\right),\label{eq:cpv}
\end{align}
where all symbols have their usual meaning. The $\mp$ sign holds for $\delta =\pm \frac{\pi}{2}$. For $\lambda = 0$, or equivalently $\theta_{13}=0$, the $U^\pm$ reduce to the standard tri-bimaximal mixing 
matrix~\cite{Harrison:2002er}.\footnote{This may be compared with~\cite[Eq.~(26)]{Stancu:1999ct} that gives an interpolating matrix with $\theta_\odot$ as a variable. In one limit the interpolating matrix gives the bimaximal mixing~\cite{Vissani:1997pa,Ahluwalia:1998xb,Barger:1998ta} and in another it yields tri-bimaximal mixing~\cite{Harrison:2002er}. }

The result~(\ref{eq:cpv}) is modified by matter effects~\cite{Wolfenstein:1977ue,Mikheev:1986gs}. Its general features are studied in detail by various authors~\cite{Gava:2008rp,Balantekin:2007es,Kneller:2009vd,Kisslinger:2012se}.
In gravitational environments the following argument suggests that one must expect a significant modification to the result~(\ref{eq:cpv}). Neutrino oscillations provide us with a set of flavour oscillation clocks. These clocks must redshift according to the general expectations of the theory of general relativity.
In gravitational environments of neutron stars the dimensionless gravitational potential is $\Phi^{NS}_{grav}\approx 0.2$ (cf. for Earth, $\Phi^{\oplus}_{grav}\approx 6.95 \times 10^{-10}$). For a given source-detector distance, and a given energy, the asymmetries $A_{CP}$ for supernovae modeling must be accordingly modified~\cite{Ahluwalia:1996ev,Ahluwalia:1998jx,Konno:1998kq,Wudka:2000rf,Mukhopadhyay:2005gb,Singh:2003sp} at the $20\%$ level, or thereabouts.

 An examination of the $U^\pm$ immediately shows that the expectation values of the $\nu_\mu$ and $\nu_\tau$ masses are identical.
To $\mathcal{O}(\lambda^2)$ the $U^-$ obtained above reproduces to King's result~\cite[Eq.~(8)]{King:2012vj} for $\delta = \pi/2$.
The presented $U^\pm$ not only accommodate the implications of the Daya Bay and RENO collaborations, but also the L/E flatness of the $e$-like event ratio seen in the atmospheric neutrino data while respecting all other known data on neutrino oscillations.

\acknowledgments
The result presented here was obtained on 10 May 2012, and was presented the next day at a MatScience Seminar. The author thanks 
Institute of Mathematical Sciences (``MatScience'', Chennai, India)   for its hospitality and for its vibrant scholarly environment.


\providecommand{\href}[2]{#2}\begingroup\raggedright
\endgroup

\end{document}